\documentstyle[emulateapj,graphicx]{article}
\begin{document}

\title{A Two-Component Explosion Model for the Giant Flare\\ and Radio Afterglow from
SGR1806-20}

\author{Z. G. Dai$^1$, X. F. Wu$^1$, X. Y. Wang$^1$, Y. F. Huang$^1$, and B. Zhang$^2$}
\affil{$^1$Department of Astronomy, Nanjing University, Nanjing
210093, China;\\ dzg@nju.edu.cn, xfwu@nju.edu.cn, xywang@nju.edu.cn, hyf@nju.edu.cn  \\
$^2$Department of Physics, University of Nevada, Las Vegas, NV 89154, USA;
bzhang@physics.unlv.edu}

\begin{abstract}
The brightest giant flare from the soft $\gamma$-ray repeater (SGR) 1806-20 was
detected on 2004 December 27. The isotropic-equivalent energy release of this
burst is at least one order of magnitude more energetic than those of the two
other SGR giant flares. Starting from about one week after the burst, a very
bright ($\sim 80$ mJy), fading radio afterglow was detected. Follow-up
observations revealed the multi-frequency light curves of the afterglow and the
temporal evolution of the source size. Here we show that these observations can
be understood in a two-component explosion model. In this model, one component
is a relativistic collimated outflow responsible for the initial giant flare
and the early afterglow, and another component is a subrelativistic wider
outflow responsible for the late afterglow. We also discuss triggering
mechanisms of these two components within the framework of the magnetar model.
\end{abstract}


\section{Introduction}

Soft $\gamma$-ray repeaters (SGRs) emit short-duration ($\sim 0.1$ seconds)
bursts of soft $\gamma$- and hard X-rays. They are believed to be magnetars,
neutron stars with long periods of a few seconds and dipole fields of $\sim
10^{15}$ Gauss (Thompson \& Duncan 1995; Woods \& Thompson 2004 for a recent
review). Only three giant flares from SGRs have been detected so far, the
brightest of which originated from SGR1806-20 on 2004 December 27 (Hurley et
al. 2005; Palmer et al. 2005; Mazets et al. 2005). The follow-up observations
of the December 27 event revealed a bright radio afterglow with the size,
proper motion, polarization, and flux at different radio frequencies all
evolving with time (Gaensler et al. 2005; Cameron et al. 2005; Gelfand et al.
2005; Taylor et al. 2005). An obvious break in the multi-frequency light curves
of the afterglow occurred around day 9 after the burst and subsequently the
flux declined rapidly as $\propto t^\alpha$ (where $\alpha\sim -2.7$) between
days 10 and 25. But the afterglow hardly faded during the period of days
25--33. In fact it underwent a slower decay. The observed temporal evolution of
the source size was $\theta(t)\propto t^{0.04\pm 0.15}$ on days 7 to 10
(Cameron et al. 2005). This was followed by a faster increase (Taylor et al.
2005; also see Granot et al. 2005). The proper motion of the afterglow on days
7--17 was negligible but subsequently became significant (Taylor et al. 2005).
In addition, the spectrum is generally consistent with a single power law index
($F_\nu\propto \nu^\beta$). Cameron et al. (2005) found $\beta=-0.62\pm 0.02$
on day 7 and then the spectrum steepened from $\beta=-0.76\pm 0.05$ on day 15
to $\beta=-0.9\pm 0.1$.

Two different models have been proposed to explain these observational results.
In Wang et al. (2005), we considered an initially relativistic energetic blast
wave from the December 27 giant flare, following the suggestion that
relativistic fireballs may occur in SGR bursts (Huang, Dai \& Lu 1998; Thompson
\& Duncan 2001). This is, in principle, the same mechanism established for GRB
afterglows (M\'esz\'aros \& Rees 1997; Sari, Piran \& Narayan 1998), i.e. the
emission from a relativistically expanding blast wave that forms when the
outflow from the SGR sweeps up the surrounding interstellar medium. To explain
the light-curve break around day 9, we invoked a broken power-law distribution
of the shock-accelerated electrons. This model appears to explain the rapid
fading of the afterglow but cannot account for the rebrightening starting on
day 25 and peaking on day 33.

Realizing this difficulty, some authors (Gaensler et al. 2005; Cameron et al.
2005; Gelfand et al. 2005; Granot et al. 2005) proposed a subrelativistic
outflow model, in which the outflow from the giant flare initially expanded in
a cavity at a constant velocity of $\sim 0.3c$ and about 7 days later it
happened to collide with a thin shell, which led to its deceleration by a
reverse shock while the ambient matter was accelerated by a forward shock. In
this model, the rapid decline of the afterglow flux was caused by adiabatic
expansion of the reversely-shocked electrons and the rebrightening was due to
initial coasting and subsequent deceleration of the outflow. This model
predicts a size $\theta(t)\propto t$ in the coasting phase, which is
inconsistent with the observations of Cameron et al. (2005). Furthermore, one
has to assume an unobserved density bump to ignite the fireball right before
the observations started. It is unclear whether this could happen and what
could be the source of the density bump

The idea of an initially subrelativistic blast wave from SGRs was suggested by
Cheng \& Wang (2003), who explained the radio afterglow light curve of the 1998
August 27 giant flare from SGR1900+14. For this event, a rebrightening could be
seen although the data were sparse (Frail et al. 1999). In this {\em Letter},
we propose a two-component explosion model to interpret the abundant
observational data for the December 27 event. In this model, one relativistic
collimated outflow is responsible for the initial giant flare and the early
afterglow, and another subrelativistic wider outflow is responsible for the
late afterglow. Our model seems to provide a unified picture in understanding
the two giant flares and their radio afterglows from SGRs. In \S 2, we discuss
why two components are needed for the giant flares and radio afterglows, and in
\S 3, we fit the observed light curves at 4.86 GHz and 8.46 GHz and the source
size as a function of time. In \S 4, we summarize our results.

\section{An Explosion with Two Components}

Within the framework of the magnetar model, Thompson \& Duncan (2001) proposed
a triggering mechanism for giant flares, in which a helical distortion of the
core magnetic field induces a large-scale cracking of the crust and a twisting
deformation of the exterior magnetic field. This mechanism is consistent with
the three observed timescales in the December 27 giant flare (Schwartz et al.
2005). It seems that the energy release for this mechanism includes two
components. First, as the crust cracks, the exterior magnetic field deforms,
which leads to a purely magnetohydrodynamic instability. Such an instability
probably produces reconnection events and induces magnetohydrodynamic waves
outside the star. The dissipated wave energy is initially locked onto the
magnetic field in a thermal photon-pair plasma, which, under its own huge
pressure, expands adiabatically along some surface magnetic lines and
eventually drives a relativistic collimated outflow because of very low baryon
loading. Second, when the core ``toroidal" magnetic field floats up to break
through the stellar crust, reconnection of the newborn surface magnetic field
will drive an explosive outflow (Klu\'zniak \& Ruderman 1998; Dai \& Lu 1998).
Because this magnetic field inevitably brings some crustal matter and the
reconnection occurs at about the stellar radius, the resultant outflow has high
baryon contamination and thus it is subrelativistic. Alternatively, even if
only one exterior magnetic-reconnection event associated with the giant flare
happens within the region whose size is less than the stellar radius above the
surface (otherwise the energy release is too small to power the observed giant
flare, because the dipole magnetic field steeply decreases with increasing
radius, i.e., $B_d\propto R^{-3}$), the outward photon-pair flow launches a
collimated outflow with very low baryon loading, as discussed above for a
relativistic component. In the meantime, the inward emission impinges on the
stellar surface, and its part is reflected back and re-emitted from a larger
solid angle with much heavier baryon loading from the surface as a
subrelativistic component.

We now discuss why two components are needed to interpret the observational
data. We consider a $\gamma$-ray energy release of the December 27 burst,
$E_0\sim 3\times 10^{46}$ ergs, near the surface of a magnetar with radius of
$R_0\sim 10^6$ cm in a time of $t_0\sim 0.2$ s, and thus the outflow luminosity
is at least $L_0\sim 10^{47}\,{\rm erg}\,{\rm s}^{-1}$, as implied by the
observed hard spike (Hurley et al. 2005). An extremely large optical depth
means a radiation-pair outflow with an initial temperature of
\begin{equation}
T_0=\left(\frac{L_0}{16\pi
R_0^2\sigma}\right)^{1/4}=210L_{0,47}^{1/4}R_{0,6}^{-1/2}\,{\rm keV},
\end{equation}
where $L_{0,47}=L_0/10^{47}\,{\rm erg}\,{\rm s}^{-1}$, $R_{0,6}=R_0/10^6{\rm
cm}$, and $\sigma$ is the Stephan-Boltzmann constant. Assuming the baryon
loading rate of the outflow, $\dot{M}$, we define a dimensionless entropy
$\eta=L_0/\dot{M}c^2$.

If $\eta>1$, this outflow will expand relativistically as the Lorentz factor
$\Gamma\propto R$ and the temperature $T\propto R^{-1}$ (Shemi \& Piran 1990).
The temperature, Lorentz factor and radiation luminosity at the photosphere
become $\Gamma_{\rm f}=\min[\eta,\eta_*]$, $T_{\rm ph}=T_0\times
\min[(\eta/\eta_*)^{8/3},1]$, and $L_{\rm ph}=L_0\times
\min[(\eta/\eta_*)^{8/3},1]$ respectively, where $\eta_*=[L_0\sigma_T/(4\pi
m_pc^3R_0)]^{1/4}\simeq 100L_{0,47}^{1/4}R_{0,6}^{-1/4}$ (M\'esz\'aros \& Rees
2000). As suggested by Nakar, Piran \& Sari (2005) and Ioka et al. (2005), the
giant flare might have arisen from the emission at the photosphere and/or
internal shocks if the outflow is variable. The observed average temperature of
the hard spike, $T_{\rm spike}=175\pm 25$ keV (Hurley et al. 2005), requires
$\eta=93L_{0,47}^{5/32}R_{0,6}^{-1/16}$.

If $\eta<1$, the outflow will also expand under its own pressure, which
includes the gas pressure and radiation pressure, $P=P_g+P_r=n_p k_BT+4\sigma
T^4/(3c)$, where $n_p=\dot{M}/(4\pi R^2m_pc)$ is the proton number density at
radius $R$. Letting $P_g=P_r$ at radius $R_0$, we define the critical baryon
loading rate, $\dot{M}_{\rm cr}=16\pi \sigma m_pR_0^2 T_0^3/(3k_B)=1.7\times
10^{29}L_{0,47}^{3/4}R_{0,6}^{1/2}\,{\rm g}\,{\rm s}^{-1}$. If the baryon
loading rate is below $\dot{M}_{\rm cr}$, the radiation pressure exceeds the
gas pressure so that the temperature decays $T\propto R^{-1}$. In this case,
the temperature and radiation luminosity at the photosphere decreases to
\begin{equation}
T_{\rm ph}=T_0R_0/R_{\rm
ph}=220L_{0,47}^{1/4}R_{0,6}^{1/2}\dot{M}_{25}^{-1}\,{\rm K},
\end{equation}
and
\begin{equation}
L_{\rm ph}=4\pi R_{\rm ph}^2\sigma T_{\rm ph}^4=2.0\times
10^{32}L_{0,47}R_{0,6}^2\dot{M}_{25}^{-2}\,{\rm erg}\,{\rm s}^{-1},
\end{equation}
where the photospheric radius $R_{\rm ph}=\sigma_T\dot{M}/(4\pi m_pc)=1.1\times
10^{13}\dot{M}_{25}$ cm with $\dot{M}_{25}=\dot{M}/10^{25}{\rm g}\,{\rm
s}^{-1}$. However, if the baryon loading rate exceeds $\dot{M}_{\rm cr}$, the
gas pressure dominates over the radiative pressure and the temperature decays
as $T\propto R^{-2}$. In this case, the temperature and radiation luminosity at
the photosphere are
\begin{equation}
T_{\rm ph}=T_0(R_0/R_{\rm ph})^2=2.0\times 10^{-5}
L_{0,47}^{1/4}R_{0,6}^{3/2}\dot{M}_{25}^{-2}\,{\rm K},
\end{equation}
and
\begin{equation}
L_{\rm ph}=1.4\times 10^4L_{0,47}R_{0,6}^6\dot{M}_{25}^{-6}\,{\rm erg}\,{\rm
s}^{-1}.
\end{equation}
Thus, the temperature and luminosity of the emission from the subrelativistic
outflow at the photosphere in both cases are much less than those of the giant
flare, showing that a subrelativistic outflow model for explaining the giant
flare can be ruled out.

Therefore, we conclude that an ultrarelativistic outflow is required by the
extremely high peak luminosity with millions of the Eddington value, hard
spectrum and rapid variability of the initial spike emission of the giant
flares. Furthermore, a relativistic jet model indeed provides a satisfactory
explanation for the observed light curve of the initial spike of the December
27 giant flare, as shown by Yamazaki et al. (2005). It is possible that such an
outflow, after emitting the hard spike, retains some amount of energy and then
drives a blast wave when sweeping into the ambient medium.

On the other hand, an obvious bump at $t_{\rm dec}\sim 30$ days in the 4.86 GHz
light curve of the radio afterglow for the December 27 event and a possible
bump at $t_{\rm dec}\sim 10$ days for the August 27 event imply that
subrelativistic outflows with initial velocities of $\beta_{{\rm nr},0} c$
could begin to be decelerated by the ambient medium with density of $n$ at
$t_{\rm dec}$. Thus we obtain the outflow mass and kinetic energy
\begin{equation}
M_{\rm nr}=1.2\times 10^{24}\beta_{{\rm nr},0}^3n_{-2}(t_{\rm dec}/10\,{\rm
days})^3\,{\rm g},
\end{equation}
and
\begin{equation}
E_{\rm nr}=5.4\times 10^{44}\beta_{{\rm nr},0}^5n_{-2}(t_{\rm dec}/10\,{\rm
days})^3\,{\rm ergs},
\end{equation}
where $n_{-2}=n/10^{-2}\,{\rm cm^{-3}}$. This outflow could also drive a
forward shock with negligible energy loss.

In short, there could have been two blast waves after the December 27 giant
flare. In the following we show that an initially relativistic shock is
responsible for the early afterglow and a nonrelativistic shock for the late
afterglow.

\section{Fitting the Afterglow Light Curve and Size}

We consider two energetic shocks after the December 27 giant flare: one
initially ultrarelativistic blast wave with an opening angle of $\theta_j$, an
isotropic-equivalent energy of $E_{\rm r}$, and the Lorentz factor $\Gamma_0\gg
1$, and another subrelativistic forward shock with an energy of $E_{\rm nr}$
and an initial velocity of $\sim 0.2c$. Both shocks expand in a uniform medium
with density of $n$. The initially relativistic blast wave will enter the
non-relativistic phase at $t_{\rm nr}=4.0E_{{\rm
r},44}^{1/3}n_{-2}^{-1/3}\,{\rm days}$ and is expected to play a dominative
role in the early radio afterglow, where $E_{{\rm r},44}=E_{\rm r}/10^{44}{\rm
ergs}$. As suggested in Wang et al. (2005), the electron energy distribution in
this blast wave is taken to be a broken power-law form with $p_1$ and $p_2$
below and above the break Lorentz factor $\gamma_b$ during the
trans-relativistic stage,
\begin{equation}
\frac{dN_e}{d\gamma_e}\propto \left\{
\begin{array}{l}
 \gamma_e^{-p_1}, \,\,\,\,\, {\rm if} \,\gamma_{\rm min}\le \gamma_e<\gamma_b \\
 \gamma_e^{-p_2}, \,\,\,\,\, {\rm if } \,\gamma_e\ge\gamma_b,
\end{array} \right.
\end{equation}
where $\gamma_{\rm min}$ is the minimum Lorentz factor. Observationally, such
an electron distribution is not only required by the spectral evolution of the
afterglow (Cameron et al. 2005) and the synchrotron spectrum of the Crab Nebula
(Amato et al. 2000) but also suggested in fitting two gamma-ray burst
afterglows (Li \& Chevalier 2001) and TeV blazars (Tavecchio, Maraschi \&
Ghisellini 1998). Theoretically, the recent particle simulation of a
relativistic two-stream instability by Dieckmann (2005) revealed a broken
power-law distribution. Wang et al. (2005) have shown that this distribution
can account for the frequency-dependent breaks of the light curves around day 9
and the steepening of the radio-band spectra with time (Cameron et al. 2005).

The initially sub-relativistic component may largely contribute to the radio
afterglow emission only at late times, as it is decelerated and the swept-up
matter accumulates. According to this scenario, the observed bump time $t\sim
30$ days corresponds to the deceleration time $t_{\rm dec}$ of this component,
so we infer $\beta_{{\rm nr},0}=R_{\rm nr}/c t_{\rm dec}=0.38(R_{\rm
nr}/3\times 10^{16}{\rm cm})(t_{\rm dec}/30\,{\rm d})^{-1}$, where $\beta_{{\rm
nr},0}$ is the initial velocity of the sub-relativistic component and $R_{\rm
nr}$ is the blast wave radius at the bump time. Then we can constrain the
isotropic-equivalent energy of this component from the deceleration time, i.e.,
$E_{\rm nr}=1.5\times 10^{44}n_{-2}(t_{\rm dec}/30\,{\rm d})^3(\beta_{{\rm
nr},0}/0.4)^5$ ergs (from equation 7). The electron energy distribution for
this component is assumed to be a single power-law one with index $p$. This is
deduced from the spectrum form of the late radio afterglow of the 1998 August
27 giant flare from SGR 1900+14.

We first carried out numerical calculations of the dynamics of each blast wave
based on Huang et al. (2000) and obtained the temporal evolution of the
source's radius. We next calculated light curves of the synchrotron radiation
from the shocked matter at different frequencies, assuming that the electron
energy distribution has the same form as equation (8) and that $\epsilon_e$ and
$\epsilon_B$ are fractions of the total energy density that go into the
electrons and magnetic field respectively. Combining the contributions from
both components, we performed numerical fitting to the light curves at 4.86 GHz
and 8.46 GHz in Figures 1 and 2 respectively. In the fitting, the physical
parameter values are taken: $\Gamma_0=100$, $E_{\rm r}=2.3\times 10^{45}\,{\rm
ergs}$, $\theta_j=0.129$, $p_1=2.2$, $p_2=3.5$, and $R_{b}\equiv
\gamma_b/\gamma_{\rm min}=20$ for the initially relativistic component, and
$E_{\rm nr}=0.75\times 10^{44}\,{\rm ergs}$, $\Gamma_{{\rm nr},0}=1.023$, and
$p=2.4$ for the initially subrelativistic component. The other parameter values
are $n=0.0363\,{\rm cm}^{-3}$, $\epsilon_{e}=0.34$, and $\epsilon_{B}=0.23$. We
take the distance to the source, $d=9.8$ kpc (Cameron et al. 2005), which is
slightly less than used in previous works. Our fitting gives the total
$\chi^2/{\rm dof}=374/(79-11)=5.5$ for 11 physical parameters.

%Similarly, we also applied the two-component model to the case of the radio
%afterglow of the 1998 August 27 giant flare from SGR 1900+14 (Frail et al.
%1999). We find that our model can also give a consistent fit to this radio
%afterglow, as shown in Figure 3. The radio afterglow flux observed about one
%week after the flare was almost entirely contributed by the sub-relativistic
%component, which is consistent with the early suggestion that this radio
%afterglow was produced by a non-relativistic outflow (Cheng \& Wang 2003). The
%relatively earlier time of the appearance of a bump in this afterglow than in
%the case of the December 27  giant flare is naturally expected to result from a
%smaller energy of the wide component for the August 27 giant flare and thus a
%shorter deceleration time. Unfortunately, the early afterglow for this giant
%flare was not seen due to the lack of sufficiently early follow-up observations
%by radio telescopes.

The evolution of the source size with time was reported for the December 27
event (Gaensler et al. 2005; Cameron et al. 2005; Granot et al. 2005; Taylor et
al. 2005). For the two-component model, the measured size at any time should be
dominated by the brighter component. In Figure 3, we present the evolution of
the source sizes of the initially relativistic and subrelativistic components,
denoted by the dashed and dotted lines, respectively. From the light curves in
Figure 2, we see that the flux density of the initially subrelativistic
component begins to dominate over that of the initially relativistic component
around day 20, so that the measured sizes should be the sizes of the initially
relativistic and subrelativistic components before and after this time
respectively. The solid line linking both components in Figure 3 indicates the
transition regime between the two phases.

\section{Conclusions}

We have proposed a two-component explosion model, in which one initially
relativistic collimated outflow can account for the spectrum of the giant flare
and the early-time broken light curves at different radio frequencies and the
slow increase of the source size, and another subrelativistic wider outflow for
the late-time rebrightening of the radio afterglow and the faster increase of
the source size. In addition, our model is consistent with the magnetar
scenario of Thompson \& Duncan (2001). In this scenario, two magnetic
reconnection events may happen near the surface and in the magnetosphere
respectively, which give rise to a relativistic collimated outflow and a
subrelativistic outflow because of different baryon loadings. Alternatively,
even if only one magnetic reconnection event associated with the giant flare
happens within a small region above the surface, the outward photon-pair flow
launches a relativistic collimated fireball while the inward flow strikes the
surface and is partially re-emitted as a subrelativistic baryonic outflow from
a larger solid angle. Finally, the relativistic component from a giant flare
like the December 27 event may be diagnosed using future high energy data, as
noted by Fan, Zhang \& Wei (2005).

%From Taylor et al. (2005), we see that the proper motion of the afterglow on
%days 7--17 after the December 27 event was negligible but subsequently became
%significant. This property cannot be imagined in any one-component explosion
%model but can be understood in our two-component explosion model: if an
%initially relativistic outflow is a jet pointing towards the line of sight,
%this outflow has no proper motion; if another initially subrelativistic
%component is an off-axis jet, its proper motion may be detected when its
%emission dominates over that from the initially relativistic outflow. Thus, the
%proper motion of the afterglow from the December 27 event may further support
%the two-component explosion picture with an initially relativistic on-axis
%outflow and a subrelativistic off-axis outflow.

{\acknowledgments We thank the referee for valuable comments. This work was
supported by the Ministry of Science and Technology of China (NKBRSF
G19990754), the Special Funds for Major State Basic Research Projects, the
FANEDD 200125, and the National Natural Science Foundation of China under
grants 10403002, 10233010 and 10221001. B. Zhang was supported by NASA
NNG04GD51G and a NASA Swift GI (Cycle 1) program.}

\begin{figure*}
\plotone{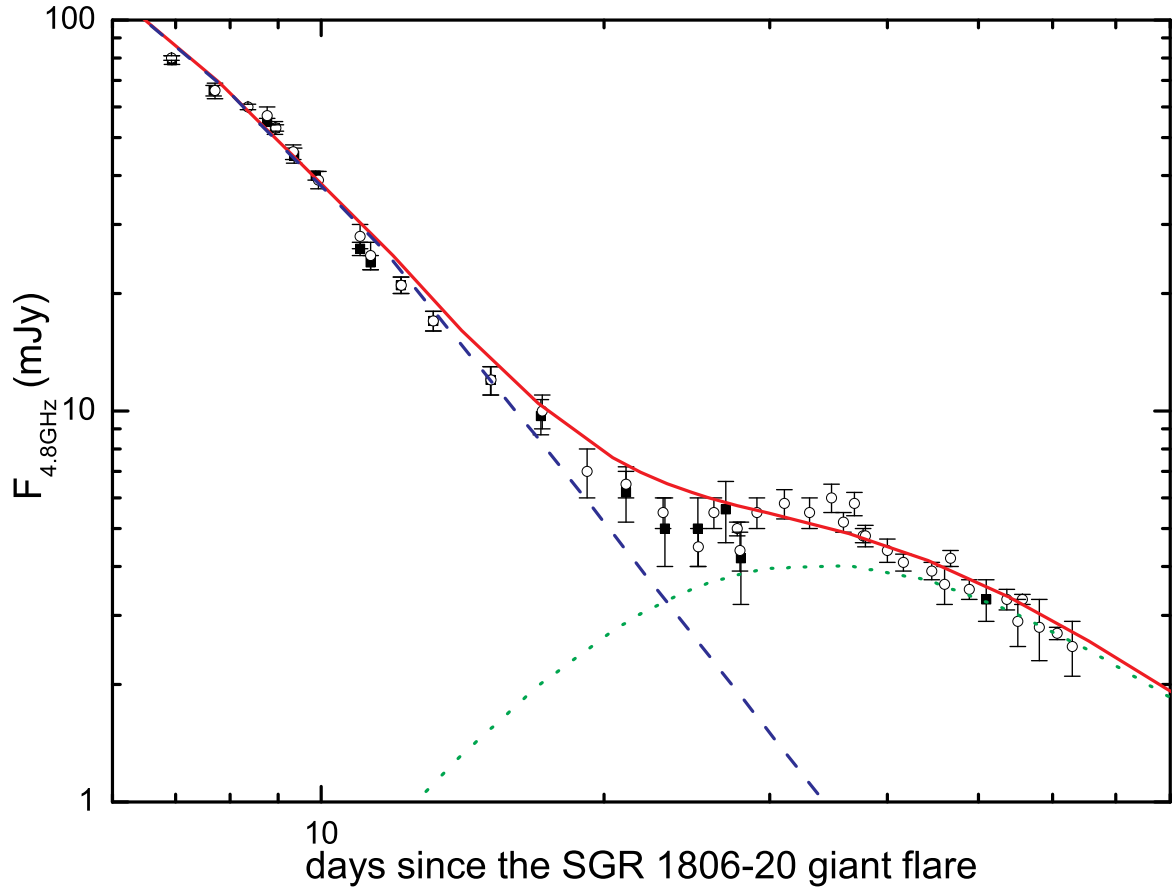} \vspace{0cm} \caption{Modelling the 4.86 GHz afterglow of the
December 27 giant flare from SGR 1806-20 in the two-component explosion model.
The physical parameter values are given in the text. The dashed and dotted
lines represent the contributions from the initially ultrarelativistic and
subrelativistic components, respectively, while the solid line is the sum of
both contributions. The data are taken from Gelfand et al. (2005, {\em open
circles}) and Cameron et al. (2005, {\em solid squares}). }
%\label{fig:rimage}}
\end{figure*}

\begin{figure*}
\plotone{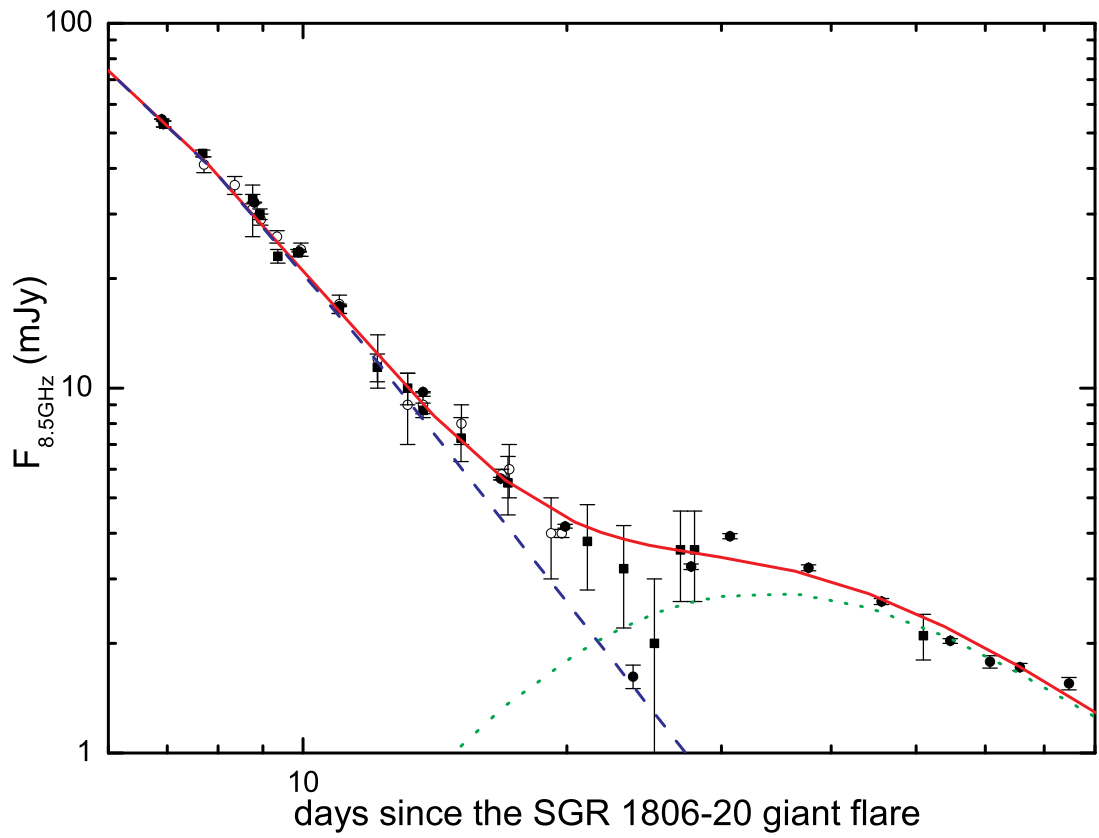} \vspace{0cm} \caption{Modelling the 8.46 GHz afterglow of the
December 27 giant flare from SGR 1806-20 in the two-component explosion model
with the same parameter values as in Figure 1. The data are taken from Cameron
et al. (2005, {\em solid squares}), Gaensler et al. (2005, {\em open circles})
and Taylor et al. (2005, {\em solid circles}). }
%\label{fig:rimage}}
\end{figure*}

%\begin{figure*}
%\plotone{f3.eps} \vspace{0cm} \caption{Modelling the 8.46 GHz afterglow of the
%1998 August  27 giant flare from SGR 1900+14 in the two-component blast wave
%model. The data are taken from Frail et al. (1999). The physical parameter
%values of both components used in the fits are $E_{\rm r}=2\times 10^{43}\,{\rm
%ergs}$, $\theta_j=0.15$, $p_1=2.2$, $p_2=3.5$, $R_{b}\equiv
%\gamma_b/\gamma_{\rm min}=10$, and $E_{\rm nr}=5.2\times 10^{42}\,{\rm ergs}$,
%$\Gamma_{{\rm nr},0}=1.054$, $p=2.65$, respectively. Other parameter values are
%$n=0.035\,{\rm cm}^{-3}$, $\epsilon_{e}=0.3$, $\epsilon_{B}=0.3$, and $d=10$
%kpc.  }
%\label{fig:rimage}}
%\end{figure*}

\begin{figure*}
\plotone{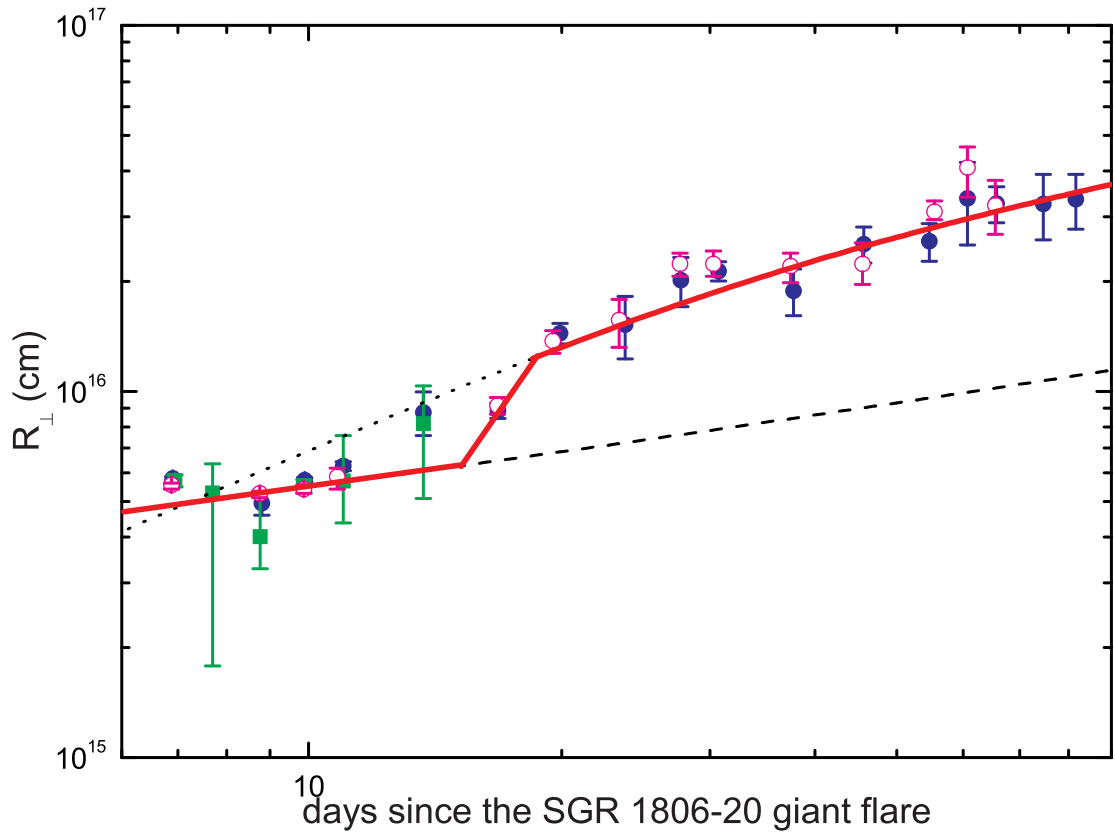} \vspace{0cm} \caption{Modelling the size evolution of the
radio afterglow of the 2004 December 27 giant flare from SGR 1806-20 in the
two-component model with the same parameter values as Figure 1. The dashed and
dotted lines represent the size evolution of the initially ultrarelativistic
and subrelativistic components, respectively. The solid line is the size of the
brighter one between both components. The data are taken from Cameron et al.
(2005, {\em solid squares}), Granot et al. (2005, {\em open circles}), and
Taylor et al. (2005, {\em solid circles}).}
%\label{fig:rimage}}
\end{figure*}

\end{document}